\documentclass[
aps,%
12pt,%
final,%
notitlepage,%
oneside,%
onecolumn,%
nobibnotes,%
nofootinbib,%
superscriptaddress,%
showpacs,%
centertags]%
{revtex4}

\usepackage[english]{babel}
\usepackage[dvips]{graphicx}
\usepackage{amssymb,amsmath,epsf}
\usepackage{epsfig}
\usepackage{wrapfig}

\def\spp{\mbox{$^3S_1^{++}$}}
\def\dpp{\mbox{$^3D_1^{++}$}}
\def\ppi{{\mbox{\boldmath$p_i$}}}
\def\ppa{{\mbox{\boldmath$p_1$}}}
\def\ppb{{\mbox{\boldmath$p_2$}}}
\def\ppp{{\mbox{\boldmath$p_p$}}}
\def\ppn{{\mbox{\boldmath$p_n$}}}
\def\pp{{\mbox{\boldmath$p$}}}
\def\p'{{\mbox{\boldmath$p'$}}}

\def\xx{{\mbox{\boldmath$x$}}}
\def\k{{\mbox{\boldmath$k$}}}

\def\s{{\mbox{\boldmath$\sigma$}}}
\def\q{{\mbox{\boldmath$q$}}}
\def\l{{\mbox{\boldmath$l$}}}
\def\r{{\mbox{\boldmath$r_1$}}}
\def\rr{{\mbox{\boldmath$r_2$}}}
\def\rrr{{\mbox{\boldmath$r_i$}}}

\def\jj{{\mbox{\boldmath$J$}}}
\def\n{{\mbox{\boldmath$\nabla$}}}

\def\zz{{\mbox{\boldmath$0$}}}
\def\re{{\mbox{\rm Re}}}

\def\ee{$E$}
\def\eepr{$E^{\prime}$}
\def\thepr{$\theta$}
\def\pn{$p_n$}

\def\thn{$\theta_n$}
\def\ppro{$p_p$}

\def\thp{$\theta_p$}
\def\thqe{$\theta_{qe}$}
\def\thpe{$\theta_{pe}$}
\def\sqs{$\sqrt{s}$}
\def\sqsmm{$\sqrt{s}-2m$}
\def\sla{S$_{I}$}
\def\slb{S$_{II}$}
\def\slc{S$_{III}$}
\def\boa{B$_{I}$}
\def\bob{B$_{II}$}
\def\boc{B$_{III}$}
\def\bod{B$_{IV}$}
\def\boe{B$_{V}$}

\begin{document}

\title{THE RELATIVISTIC IMPULSE APPROXIMATION FOR THE
EXCLUSIVE ELECTRODISINTEGRATION OF THE DEUTERON}

\author{\firstname{S. G.} \surname{Bondarenko}}
\noaffiliation
\author{\firstname{V. V.} \surname{Burov}}
\noaffiliation
\author{\firstname{E. P.} \surname{Rogochaya}}
\affiliation{Joint Institute for Nuclear Research, 141980,
Dubna, Moscow region, Russia}
\author{\firstname{A. A.} \surname{Goy}}
\affiliation{Far Eastern National University, 690950, Vladivostok,
Russia}

\begin{abstract}
The electrodisintegration of the deuteron in the frame of the
Bethe-Salpeter approach with a separable kernel of the
nucleon-nucleon interaction is considered. This conception keeps
the covariance of a description of the process. A comparison of
relativistic and nonrelativistic calculations is presented. The
factorization of the cross section of the reaction in the impulse
approximation is obtained by analytical calculations. It is shown
that the photon-neutron interaction plays an important role.
\end{abstract}

\pacs{25.30.Dh}

\maketitle

\section{Introduction}
\label{intro}

Study of static and dynamic electromagnetic properties of light
nuclei enables more deeply to understand a nature of strong
interactions and, in particular, nucleon-nucleon (NN)
interactions. Deuteron as a two nucleon system is a simplest
object for the NN interaction investigation.

First experimental data got in 1960-th \cite{crossaux} were
satisfactorily described in the framework of the nonrelativistic
formalism. But when the precision of experiments and energies of
particles taking part in the reaction increased nonrelativistic
models failed. So it became clear that relativistic effects (which
{\em a priori} are very important at large transfer momenta) are
needed to include in the consideration.

Besides the extraction from experiments with light nuclei of the
information about a structure of bounded nucleons requires to take
into account relativistic kinematics of the reaction and dynamics
of interaction. So the construction of a covariant approach and
detailed analysis of relativistic effects in electromagnetic
reactions with light nuclei are very important. It can not be
performed in the nonrelativistic picture. This understanding is an
additional reason for construction of the relativistic approach.

The electrodisintegration of the deuteron at the threshold has
been of interest of an investigation for a long time
\cite{arenh3}-\cite{gakh}. The reason is that the
electrodisintegration is an essential instrument for study a
structure of a two-nucleon system. First of all it is an
electromagnetic structure. The deuteron has been used as a neutron
target to get the information about neutron electromagnetic form
factors. During last 20 years it has also been used to receive
constraints on available realistic NN potentials. Analyzing of the
electrodisintegration process we can clarify the role of
non-nucleonic degrees of freedom.
The deuteron is one of convenient candidates because complete
calculations can in principle be performed.

The experimental results on the differential cross section derived
from $(ed \rightarrow enp)$ reaction are available up to a
momentum transfer of about 1 GeV \cite{bernheim}-\cite{boden}.
This situation is very good for investigation of the deuteron
structure at short distances with the allowance for some exotic
effects which have not been earlier important. First of all these
are the quark degrees of freedom (see \cite{kiss},
\cite{our-quarks}, for instance) but formerly it is necessary to
take into account relativistic effects.

There are several approaches for the theoretical description of
the deuteron and, in particular, the deuteron break-up reaction.
One group uses a numerical solving of a relativistic wave equation
for NN system based on the relativistic one-boson-exchange (OBE)
model (see, for instance, \cite{gross1}-\cite{gross}). Other one
used a simple phenomenological approach by adding lowest-order
relativistic corrections to the non-relativistic one-body current
and including the kinematic wave function boost \cite{arenh3} or
the covariant models based on the direct evaluation of those
Feynmann diagrams which give the dominant contributions in the
quasi-free region \cite{adam}. These is also the approach with the
using the Paris potential \cite{shebeko}.

The Bethe-Salpeter (BS) approach~\cite{BS} can give a possibility to
consider relativistic effects by consistent way \cite{burov}. In
the paper the deuteron electrodisintegration within the covariant
BS approach with the separable Graz II interaction kernel is
presented. The exclusive differential cross section is calculated
in the relativistic impulse approximation (RIA) with plane waves
in final $np$-state.

The paper is organized as follows. In section~\ref{kinem} the
relativistic kinematics of the reaction is considered, formula for
the cross section is presented. Sections~\ref{bsad}
and~\ref{bsanp} are devoted to the BS amplitude of the deuteron
and $np$-pair. The relativistic consideration of the hadron
current in the BS formalism is defined in section~\ref{hemc}. In
section~\ref{nrf} the nonrelativistic approach is taken up.
Factorization of the cross section is discussed in
section~\ref{factor}. Then the results of our numerical
calculations are presented in section~\ref{results}. Finally we
review the results and outline further plans.

\section{Cross section and kinematics}
\label{kinem}

\begin{figure}[ht]
\begin{center}
\includegraphics[width=65mm]{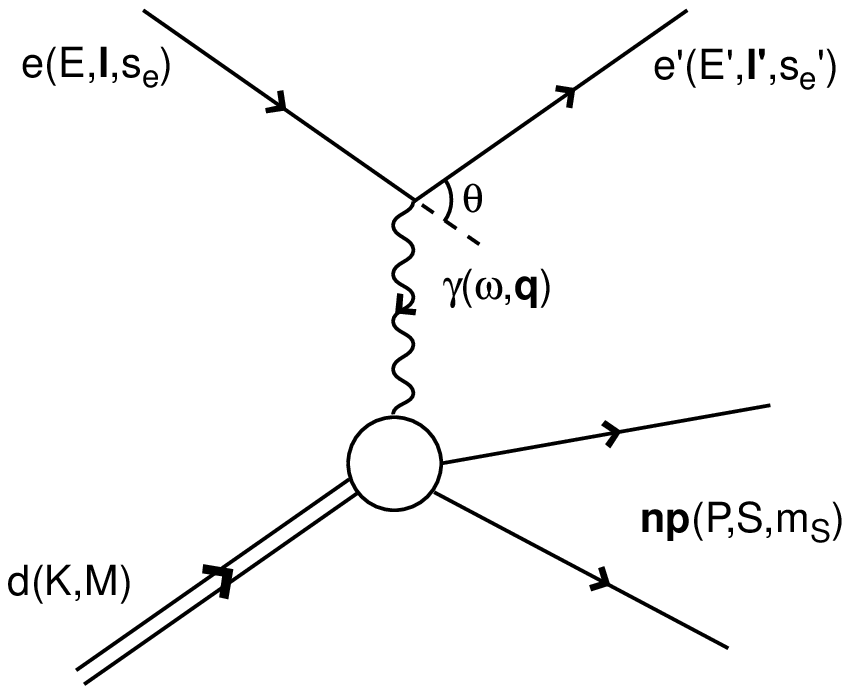}
\caption{\label{OPA} One photon approximation.}
\end{center}
\end{figure}

Let us consider the relativistic kinematics of the exclusive
electrodisintegration of the deuteron.  The initial electron
$l=(E,\l)$ collides with the deuteron in rest frame $K=(M_d,\zz)$
($M_d$ is a mass of the deuteron). There are three particles in
the final state, i.e. electron $l'=(E',\l')$ and pair of proton
and neutron. Neglecting an electron mass in one photon
approximation we can express a squared momentum of the virtual
photon $q=(\omega,\q)$ via electron scattering angle $\theta$
\begin{eqnarray}
q^2=-Q^2=(l-l^{\prime})^2=\omega^2-\q^2=-4|\l||\l'|\sin^2\frac{\theta}{2}.
\end{eqnarray}
$np$-pair is described by the invariant mass $s=P^2=(p_p+p_n)^2$
which can be written through components of photon four-impulse:
\begin{eqnarray}
s=M_d^2+2M_d\omega+q^2.
\end{eqnarray}
Lorentz invariant matrix element of the reaction (see
Fig.~\ref{OPA}) can be written as a product of lepton and hadron
currents
\begin{eqnarray}
M_{fi}=-\imath e^2(2\pi)^4\delta^{(4)}(K-P+q)
<l',s_e'|j^{\mu}|l,s_e>
\frac{1}{q^2}<np:(P,Sm_S)|J_\mu|d:(K,M)>,
\end{eqnarray}
where $<l',s_e'|j^{\mu}|l,s_e> = \bar
u(l',s_e')\gamma^{\mu}u(l,s_e)$ is an electromagnetic current
(EM). Dirac spinor $u(l,s_e)$ ($\bar u(l',s_e')$) describes the
initial (final) electron with spin projection $s_e$ ($s_e'$).
The hadron current
$<np:(P,Sm_S)|J_\mu|d:(K,M)>$ is a transition matrix element from
the initial deuteron $|d:(K,M)>$ with total momentum $K$,
projection $M$ to the final $np$-pair $|np:(P,Sm_S)>$\ with total
momentum $P$ and spin $S$, projection $m_S$. 
The unpolarized cross section of the electrodisintegration of the
deuteron can be easily written as a production of electron
$l^{\mu\nu}$ and hadron $W_{\mu\nu}$ parts:
\begin{eqnarray}
\frac{d^5\sigma}{dE^{\prime}d\Omega^{\prime}d\Omega_\pp}=
\frac{\alpha^2}{8M_d(2\pi)^3} \frac{|\l'|}{|\l|}\frac{\sqrt
s}{q^4}{R}\, l^{\mu\nu}W_{\mu\nu} \label{crosssection1}
\end{eqnarray}
with some kinematical factor $R$ which connects the final proton
angle in the center-of-mass system (C.M.S.) (where the $np$-pair
is rest) with the same in the laboratory system (L.S.) $\pp$:
\begin{eqnarray}
R=\frac{\pp^2}{\sqrt{1+\eta}|\pp|-e_\pp\sqrt{\eta}
\cos\theta_{\pp}},
\end{eqnarray}
$e_\pp=\sqrt{\pp^2+m^2}$, $\theta_\pp$ is an angle between the
final proton and $Z$-axis, $m$ is a nucleon mass, and
$\eta=\q^2/s$.

Tensor of unpolarized leptons in (\ref{crosssection1}) is
expressed in a standard form
\begin{eqnarray}
l_{\mu\nu}= \frac12 \sum_{s_e\,s_e^{\prime}}
<l^{\prime},s_e^{\prime}|{{j_{\mu}}^{\dag}}|l,s_e>
<l,s_e|j_{\nu}|l^{\prime},s_e^{\prime}> = 2(l^{\prime}_{\mu}
l_\nu+l^{\prime}_{\nu} l_{\mu} )+g_{\mu\nu}q^2
\end{eqnarray}
and hadron tensor can be written as a production of hadron
currents with averaging-out by the deuteron angle momentum
\begin{eqnarray}
W_{\mu\nu}= \frac13 \sum_{M S m_s}<d:(K,M)|{J_{\mu}}^{\dag}|np:(P,Sm_S)>
<np:(P,Sm_S)|J_{\nu}|d:(K,M)>.
\end{eqnarray}

In most cases in order to average on initial and sum on final
states it is convenient to introduce a helicity tensor which can
be directly connected with structure functions (see, for example
\cite{arenh3},\cite{gross},\cite{gakh}). These quantities allow to
calculate polarization and asymmetry observables easily and will
be necessary in future (we didn't calculate them in this work).
Keeping in mind the Hermitian properties of the lepton and the
hadron tensors the cross section can be rewritten as
\begin{eqnarray}
\frac{d^5\sigma}{dE^{\prime}d\Omega^{\prime}d\Omega_\pp}=
\frac{\sigma_{Mott}}{8M_d(2\pi)^3}{\sqrt s}{R}
\hskip 70mm
\nonumber\\
\times\left[ l^0_{00}W_{00}+l^0_{++}(W_{++}+W_{--})+ l^0_{+-}2 \re
W_{+-}-l^0_{0+}2 \re (W_{0+}-W_{0-})\right], \label{crosssection2}
\end{eqnarray}
where $\sigma_{Mott} = (\alpha \cos\frac{\theta}{2}/
2E\sin^2\frac{\theta}{2})^2$ is Mott cross section for point-like particles
and
\begin{eqnarray}
l_{00}^0=\frac{Q^2}{\q^2},\quad l_{0+}^0=\frac{Q}{|\q|\sqrt
2}\sqrt{\frac{Q^2}{\q^2}+\textmd{tg}^2 \frac{\theta}{2}},\quad
l_{++}^0=\frac{1}{2}\textmd{tg}^2\frac{\theta}{2}+\frac{Q^2}{4\q^2},\quad
l_{+-}^0=-\frac{Q^2}{2\q^2}.
\end{eqnarray}
So the calculation of the cross section (\ref{crosssection2})
comes to the calculation of the hadron tensor $W_{\mu\nu}$ which
describes the NN interaction in the deuteron and is a main subject
of our investigation.

\section{Bethe-Salpeter amplitude of the deuteron}
\label{bsad}

In the Bethe-Salpeter approach the deuteron as a system of two
bounded particles can be described by the amplitude (BSA)
$\Phi_{M}(k;K)$ which satisfies the Bethe-Salpeter equation. For
the details of using formalism we refer to~\cite{burov}. Here we
present the BS vertex function for the $\spp-\dpp$ waves used in
the calculations (for a rest deuteron $K=(M_d,\zz)$):
\begin{eqnarray}
\Gamma_{M}(k_0,\k)= \Gamma_{\spp}(\k) g_{\spp}(k_0,|\k|)
+ \Gamma_{\dpp}(\k) g_{\dpp}(k_0,|\k|),
\label{sdwaves}
\end{eqnarray}
where the spin-angular parts can be written as
\begin{eqnarray}
\Gamma_{a}(\k) = \frac{1}{\sqrt{8\pi} 2 e_\k (m+e_\k)}
(m+q_2\gamma) \frac{1+\gamma_0}{2}
{\cal G}_{a M}(\k)
(m-q_1\gamma),
\label{span-d}
\end{eqnarray}
with
\begin{eqnarray}
{\cal G}_{a M}(\k) =
\left\{
\begin{array}{lc}
\xi_{M}\gamma, & a = \spp \\
-\xi_{M}\gamma+3/(2\sqrt{2}\k^2) (q_1 \xi_M)
(q_1\gamma - q_2\gamma), & a = \dpp
\end{array}
\right.
\nonumber
\end{eqnarray}
The on-mass-shell four-vectors $q_1, q_2$ are connected with the
relative momentum $k=(k_0,\k)$ as $q_1 = (e_\k,\k), q_2 =
(e_\k,-\k)$.

For radial parts of the BS vertex function $g_a(k_0,|\k|)$ we use
the covariant Graz II kernel of $NN$-interaction.

\section{Bethe-Salpeter amplitude of the $np$-pair}
\label{bsanp}

The BSA of the $np$-pair $\chi_{Sm_s}(p;\hat p,P)$ satisfies the
inhomogeneous equation
\begin{eqnarray}
\chi_{Sm_s}(p;\hat p,P) = \chi_{Sm_s}^{(0)}(p;\hat p,P) +
\dfrac{\imath}{4\pi^3} S_2(p;P)\int\!d^4k\,
V(p,k;P)\chi_{Sm_s}(k;p,P), \label{t05c}
\end{eqnarray}
with $\hat p\cdot P = 0$ and $\hat p^2 = -s/4+m^2$ putting the
outgoing particles onto the mass shell and $S_2(p;P)=
S^{(1)}(P/2+p)S^{(2)}(P/2-p)$. The first term
$\chi_{Sm_s}^{(0)}(p;\hat p,P)$ in Eq.~\eqref{t05c} is an
amplitude which describes the free motion of two nucleons:
\begin{eqnarray}
\chi_{Sm_s}^{(0)}(p;\hat p,P) = (2\pi)^4 \delta^{(4)}(p-\hat p)
\chi_{Sm_s}^{(0)}(\hat p,P),
\label{pwa}
\end{eqnarray}
where
\begin{eqnarray}
\chi_{Sm_s}^{(0)}(\hat p,P) =
\sum_{m_1m_2} C_{\frac12 m_1 \frac12 m_2}^{Sm_S}
u_{m_1}^{(1)}(\ppa) u_{m_2}^{(2)}(\ppb),
\label{planew}
\end{eqnarray}
here four-momenta $p_1, p_2$ are in on-mass-shell form $p_1 =
(e_p,\ppp), p_2 = (e_n,\ppn)$.

Neglecting second part in Eq.~(\ref{t05c}) we introduce the plane
wave approximation (PWA). Thus the final nucleons are described by
plane waves Eq.~(\ref{planew}) and we can write in the matrix
representation for the conjugated function
\begin{eqnarray}
{\bar\chi}_{Sm_s}^{(0)}(\hat p,P) =
\frac1{\sqrt{2}}\frac1{\sqrt{2e_n(m+e_n)}}\frac1{\sqrt{2e_p(m+e_p)}}
(m-p_2\gamma)
\left[
\begin{array}{c}
-\gamma_5\\
\xi^*_{m_s}\gamma
\end{array}
\right]
\frac{1+\gamma_0}{2}(m+p_1\gamma).
\label{span-np}
\end{eqnarray}
First (second) line in the brackets stands for $S=0$ ($S=1$) case
and four-vector $\xi_{m_s}$ describes the polarization of the
$np$-pair.

\section{Hadron electromagnetic current}
\label{hemc}

If the deuteron wave function is known we are able to write the
matrix element of the hadron electromagnetic current with the BS
amplitude using Mandelstam technique \cite{mandelstam}
\begin{eqnarray}
<np:(P,Sm_S)|J_\mu|d:(K,M)>=
 \hskip 70mm
 \nonumber\\
\imath\int \frac{d^4p}{(2\pi)^4}\,\frac{d^4k}{(2\pi)^4}\,
\bar\chi_{Sm_s}(p;\hat p,P)\,\Lambda_{\mu}(p,k;P,K)\,\Phi_M(k;K).
\end{eqnarray}

We consider the process of the electrodisintegration of the
deuteron in RIA (see Fig. \ref{kinematics}).
In our further calculation only one-body currents are taken into
account
\begin{eqnarray}
\Lambda^{[1]}_{\mu}(p,k;P,K) = \imath(2\pi)^4 \left\{
\delta^{(4)}(p-k-\frac{q}{2})
\Gamma^{(1)}_\mu\left(\frac{P}{2}+p,\frac{K}{2}+k\right)
{S^{(2)}\left(\frac{P}{2}-p\right)}^{-1} \right. \hskip 5mm
\nonumber\\
\left. +\delta^{(4)}(p-k+\frac{q}{2})
\Gamma^{(2)}_\mu\left(\frac{P}{2}-p,\frac{K}{2}-k\right)
{S^{(1)}\left(\frac{P}{2}+p\right)}^{-1} \right\}
\end{eqnarray}
(here the total and relative momenta are introduced: $P=p_1+p_2,
K=k_1+k_2, k=\frac{1}{2}(k_1-k_2), p=\frac{1}{2}(p_1-p_2)$).

In this case the matrix element of the hadron current has the
following form
\begin{eqnarray}
<np:(P,Sm_S)|J_\mu|d:(K,M)>= \imath \sum_{\ell=1,2} \int
\frac{d^4p}{(2\pi)^4}\ \bar\chi_{Sm_S}(p;P)
\nonumber\\
\Gamma_\mu^{(\ell)}(q)\
S^{(\ell)}\left(\frac{P}{2}-(-1)^{\ell}p-q\right)\
\Gamma_{M}\left(p+(-1)^{\ell}\frac{q}{2};K\right).
\label{emcurrent1}
\end{eqnarray}
Note that $\gamma NN$- vertex was taken on-mass-shell
\begin{eqnarray}
\Gamma_\mu^{(\ell)}(p',p)\longrightarrow\Gamma^{(\ell)}_\mu(q)=\gamma_\mu
F^{(\ell)}_1(q^2) -\frac{1}{4m}\left(\gamma_\mu q\gamma-q\gamma
\gamma_\mu\right)F^{(\ell)}_2(q^2).
\label{gammaNN}
\end{eqnarray}
Here $F^{(\ell)}_1$ ($F^{(\ell)}_2$) - Dirac (Pauli) form factor
of the nucleon which obeys the next normalization conditions
\begin{eqnarray}
F^{(1)}_1(0) = 1,\qquad F^{(1)}_2(0) = \varkappa_p,
\nonumber\\
F^{(2)}_1(0) = 0,\qquad F^{(2)}_2(0) = \varkappa_n,
\end{eqnarray}
$\varkappa_p$ ($\varkappa_n$) is the anomalous proton (neutron)
magnetic moment.

Using (\ref{pwa}) and integrating (\ref{emcurrent1}) over $p$ we
obtain our basic PWA RIA (see Fig.~\ref{pwaria}) expression for
the hadron current
\begin{eqnarray}
<np:(P,Sm_S)|J_\mu|d:(K,M)>=
\hskip 70mm \nonumber\\
\imath \sum_{\ell=1,2}
\bar\chi^{(0)}_{Sm_S}(\hat p,P) \Gamma_\mu^{(\ell)}(q)
S^{(\ell)}\left(\frac{K}{2}-\hat p-(-1)^\ell\frac{q}{2}\right)
\Gamma_{M} \left(\hat p+(-1)^\ell\frac{q}{2};K \right).
\label{emcurrent2}
\end{eqnarray}
It has very simple form. And to get it we should just perform the
analytical calculation of the trace. For this purpose we use the REDUCE
system.

\section{Nonrelativistic formalism}
\label{nrf}

The results obtained within the relativistic frame are compared
with nonrelativistic calculations. In this case the
nonrelativistic Graz II interaction kernel \cite{mathe} for the
deuteron consideration is used. The matrix element for the hadron
current
\begin{eqnarray}
<Sm_s|j_\mu(q)|1M>=\int dx~e^{-\imath qx}<Sm_s|j_\mu(x)|1M>
\end{eqnarray}
is constructed according to a standard nonrelativistic
description with using the translational invariance property
\begin{eqnarray}
j_\mu(x)=e^{\imath Px}j_\mu(0)e^{-\imath Px}.
\end{eqnarray}
Here $P$ is a total deuteron impulse. The four-dimensional current
is defined by the expression
\begin{eqnarray}
\hat
J_\mu(0,\r,\rr)=\left(\hat\rho(0,\r,\rr)),\hat\jj(0,\r,\rr)\right)
\end{eqnarray}
where
\begin{eqnarray}
\hat\rho(\xx,\r,\rr)=\sum_{i=1,2}e_i\delta(\xx-\rrr),
\hskip60mm
\nonumber\\
\hat\jj(\xx,\r,\rr)=
\sum_{i=1,2}\left\{\frac{e_i}{2m_i\imath}\left[
\psi_\beta^*(\n_i\psi_\alpha)-(\n_i\psi_\beta^*)\psi_\alpha\right]
+\frac{e_i}{2m_i}\varkappa_i\n_i\times\left[\psi_\beta^*\s_i\psi_\alpha\right]
\right\}\delta(\xx-\rrr).
\nonumber
\end{eqnarray}
Here summation over nucleons $i$ in the deuteron is performed,
$\varkappa_i$ is the anomalous magnetic moment of the corresponding
nucleon. The first term in $\hat\jj$ corresponds to an electric
transition, the second one - to a magnetic transition. For the
deuteron we use the following wave function
\begin{eqnarray}
\psi_{1M}(\pp)=\frac{1}{\sqrt{4\pi}}\chi_{1M}u(p)+\sum_{m\mu}
C_{2m1\mu}^{1M}{\cal Y }_{2m}(\hat\pp)\chi_{1\mu}w(p).
\end{eqnarray}
After some transformations in PWA (\ref{pwa}) we get
the hadron matrix elements:\\
for the charge operator:
\begin{eqnarray}
<Sm_s|\hat\rho(0)|1M>=
\delta_{S1}
\sum_{i=1,2}
F^{(i)}_{1}(q^2) \left[\frac{1}{\sqrt{4\pi}}\delta_{Mm_s}
u(|\ppi|)+\sum_mC_{2m1m_s}^{1M}{\cal Y}_{2m}(\widehat{\ppi})
w(|\ppi|)\right],
\nonumber
\end{eqnarray}
for the electric component of current operator:
\begin{eqnarray}
<Sm_s|\hat\jj_{el}(0)|1M>=\frac{\hat \pp}{m}
\delta_{S1}
\sum_{i=1,2}
F^{(i)}_{1}(q^2) \left[\frac{1}{\sqrt{4\pi}}\delta_{Mm_s}
u(|\ppi|)+\sum_mC_{2m1m_s}^{1M}{\cal Y}_{2m}(\widehat{\ppi})
w(|\ppi|)\right].
\nonumber
\end{eqnarray}

The most cumbrous component is the magnetic component of current
operator. Therefore we separate cases for different values of
final pair spin moment $S$:
\begin{eqnarray}
<00|\hat J_{mag,\ {\lambda}}(0)|1M>=
\frac{-q_z}{m\sqrt 2}(-1)^{\lambda+i}C_{1{\lambda}10}^{1{\lambda}}
\sum_{i=1,2}
G^{(i)}_{M}(q^2)
 \times\hskip 20mm
 \nonumber\\
\left[\frac{\delta_{M,-{\lambda}}}{\sqrt{4\pi}}
u(|\ppi|)+\sum_{m'}C_{2m'1-{\lambda}}^{1M}{\cal
Y}_{2m'}\left(\widehat{\ppi}\right)w(|\ppi|)\right],
\nonumber
\end{eqnarray}
\begin{eqnarray}
<1m_s|\hat J_{mag,\ {\lambda}}(0)|1M>=
\frac{q_z}{m}C_{1{\lambda}10}^{1{\lambda}}
\sum_{i=1,2}
G^{(i)}_{M}(q^2)
 \times\hskip 20mm
 \nonumber\\
\left[\frac{1}{\sqrt{4\pi}}C_{1M1{\lambda}}^{1m_s}
u(|\ppi|)+
\sum_{m'\mu}C_{1\mu
1{\lambda}}^{1m_s}C_{2m'1\mu}^{1M} {\cal
Y}_{2m'}\left(\widehat{\ppi}\right)w(|\ppi|)
\right],
\nonumber
\end{eqnarray}
where ${\lambda}$ is a cyclic component of the vector $\hat\jj$ and
$\ppa = \ppn$, $\ppb = \ppp$.
Using the presented expressions we construct the hadron tensor and
calculate the cross section in the nonrelativistic case.

\section{Factorization of the cross section}
\label{factor}

Let us consider the electrodisintegration of the deuteron
supposing that an initial lepton collides only with the proton in
the deuteron and the neutron is a spectator. In this case the
cross section is factorized on two parts, one is connected with
the contribution of the neutron as a spectator and another with a
proton contribution, the latter does not have interference terms
between the $S$ and $D$ deuteron states.

\subsection{Nonrelativistic case}

The amplitude of the process can formally be presented
as a production
\begin{eqnarray}
{\cal M}=\chi^+_{m_1}\chi^+_{m_2}\hat O\Psi_M,
\end{eqnarray}
where spinors $\chi^+_{m_1},~\chi^+_{m_2}$ describe the outgoing
$np$-pair, $\hat O$ corresponds to the interaction vertex,
$\Psi_M$ is a wave function of the deuteron. Let us note that the
vertex $\hat O$ stands in general for any one-particle interaction
but in this paper describes $\gamma N N$-vertex.

Inserting into this expression a complete set of pair states we
can get the matrix element
\begin{eqnarray}
{\cal M}_{\mu}=\sum_{m_1'}(\chi^+_{m_1}\hat
O_{\mu}\chi_{m_1'})\chi^+_{m_1'}\chi^+_{m_2}\Psi_M
\end{eqnarray}
which can be used to derive the hadron tensor. After evident
transformations it can be written as
\begin{eqnarray}
W_{\mu\nu}= \frac13 \sum_{m_1 m_2 M} {\cal M}_{\mu}{\cal M}_{\nu}
&=& \frac{1}{3}\sum_{m_1m_2M}\left|\sum_{m_1'}\left[\chi^+_{m_1}
\hat O\chi_{m_1'}\right]\left[\chi^+_{m_1'} \chi^+_{m_2}
\Psi_M\right]\right|^2\nonumber\\
&=&\frac{1}{3}\sum_{m_1m_2M} \sum_{m_1'm_1''}\left[
\chi_{m_1}^+\hat O\chi_{m_1'}\right]^2\left[\chi^+_{m_1'}
\chi_{m_2}^+\Psi_M\Psi^+_M\chi_{m_2}\chi_{m_1''}\right].
\nonumber
\end{eqnarray}
Introducing the partial-wave decomposition for the deuteron:
$$\Psi_M =\sum_{lm s_1 s_2 s} C_{lm 1 s}^{1M}
C_{\frac{1}{2}s_1\frac{1}{2}s_2}^{1 s}Y_{lm}
\chi_{s_1}\chi_{s_2}u_l$$ and transforming the second term with
the help of orthogonalization properties of the spinor $\chi$ and
some relations for Clebsh-Gordan coefficients we can finally
obtain the factorized expression
\begin{eqnarray}
W_{\mu\nu} = \frac{1}{8\pi}|A_{\mu}A^*_{\nu}| \sum_l|u_l|^2,
\end{eqnarray}
where $A_{\mu}=\chi_{m_1}^+\hat O_{\mu}\chi_{m_1'}$ is a one-body
interaction part and $l$ counts partial states of the deuteron.

Thus the double factorization is seen.  The cross section is
proportional to the sum of squared partial radial parts of the
deuteron wave function multiplied by factorized interacting proton
part.

\subsection{Relativistic case}

In the relativistic case the matrix element of the deuteron
electrodisintegration can be written schematically in the
following form
\begin{eqnarray}
{\cal M}=\Psi_{pair}\otimes\hat O\otimes S\otimes\Gamma_M,
\end{eqnarray}
where $\Psi_{pair}$ is the wave function of the $np$-pair,
$\hat O$ is a vertex of interaction,
$S$ is a propagator of the nucleon,
$\Gamma_M$ is the vertex function of the deuteron.

\begin{figure}[ht]
\begin{center}
\includegraphics[width=65mm]{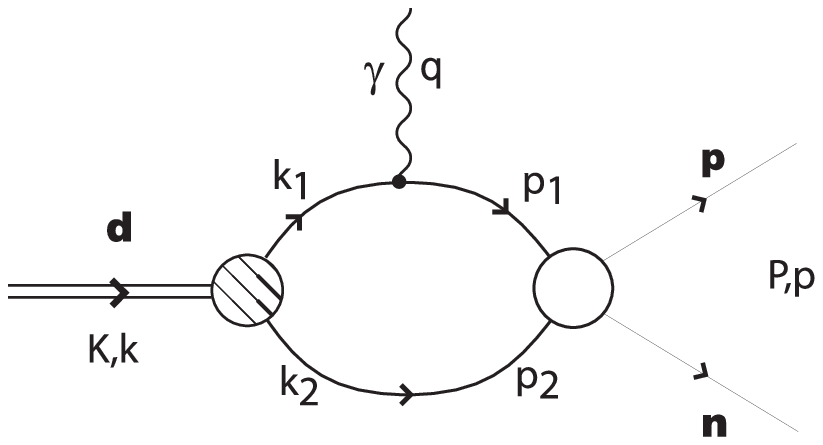}~~~~~~~~~~~~~~~~~
\includegraphics[width=65mm]{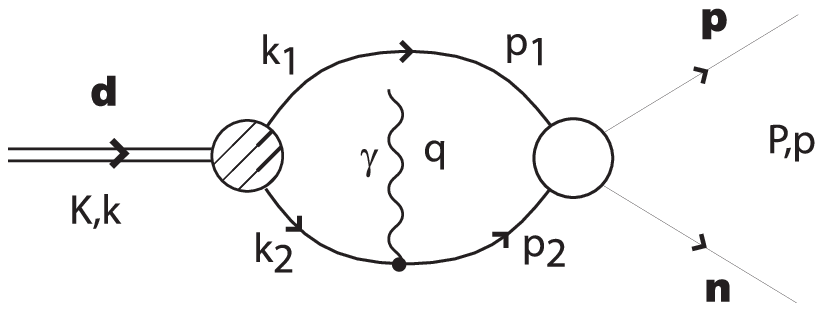}
\caption{\label{kinematics} Relativistic impulse
approximation.}
\end{center}
\end{figure}

\begin{figure}[ht]
\begin{center}
\includegraphics[width=120mm]{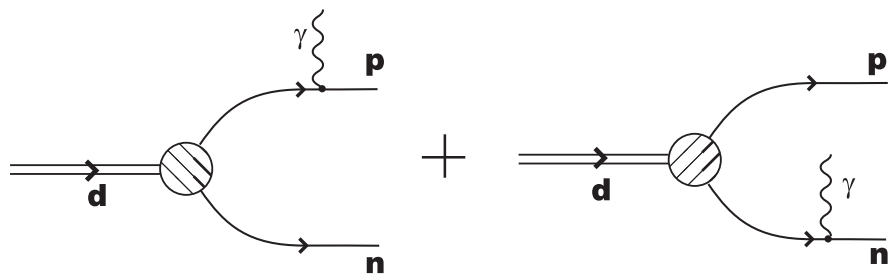}
\caption{\label{pwaria} Plane wave approximation.}
\end{center}
\end{figure}

Introducing the partial-wave decomposition of the deuteron vertex
function in the L.S., considering the only proton interacting with
a virtual photon, and supposing the PWA for the final $np$-pair we
can present the comprehensive expression in the following form
\begin{eqnarray}
{\cal M}_{\mu}&=&\sum_{s_1s_2}C_{\frac{1}{2}s_1\frac{1}{2}
s_2}^{Sm_s} \bar u_{s_1}^{(1)}(\pp_1)\bar u_{s_2}^{(2)}(\pp_2)
\Gamma^{(1)}_\mu(q)
S_{1\,+}(\k_1) \sum_{m_1'}u_{m_1'}^{(1)}(\k_1)
\bar u^{(1)}_{m_1'}(\k_1)
\nonumber\\
&&\times\sum_{m_1m_2lmsm_s}C_{lmsm_s}^{1M}
C_{\frac{1}{2}m_1\frac{1}{2}m_2}^{sm_s}u^{(1)}_{m_1}(-\k_1)
u_{m_2}^{(2)}(-\k_2)Y_{lm}(\hat\k) g_{l}(k_0,|\k|),
\nonumber
\end{eqnarray}
where $S_{1\,+}(\k_1) = 1/(k_{10}-e_{\k_1})$ and
$\Gamma^{(1)}_\mu(q)$ vertex is described by Eq.~(\ref{gammaNN}).
As it was assumed above only $^3S_1^+,~^3D_1^+$-states are taken
into account. Using orthogonalization properties of the bispinors
and some relations for Clebsh-Gordan coefficients we can write
\begin{eqnarray}
{\cal M}_{\mu} = \sum_{s_1s_2m_1lmsm_s}
C_{\frac{1}{2}s_1\frac{1}{2}s_2}^{Sm_s}C_{lmsm_s}^{1M}
C_{\frac{1}{2}m_1\frac{1}{2}s_2}^{sm_s}Y_{lm}(\hat\k)
g_{l}(k_0,|\k|) S_{1\,+}(\k_1)
A^{(1)}_{\mu}(s_1,\pp_1;m_1,\k_1),
\nonumber
\end{eqnarray}
with
\begin{eqnarray}
A^{(1)}_{\mu}(s_1,\pp_1;m_1,\k_1)=\bar u^{(1)}_{s_1}(\pp_1)
\Gamma^{(1)}_\mu(q) u_{m_1}^{(1)}(\k_1)
\end{eqnarray}
is a one-body photon-proton interaction part. Now we can
derive the hadron tensor
\begin{eqnarray}
W_{\mu\nu}=\frac{1}{3}\sum_{MSm_s}{\cal M}_{\mu}{\cal M}_{\nu}.
\nonumber
\end{eqnarray}
Using once more properties of the Clebsh-Gordan coefficients and
Dirac spinors we obtain the expression
\begin{eqnarray}
W_{\mu\nu}= C_d\ Sp \left\{(p_1\gamma+m) \Gamma^{(1)}_\mu(q)
(k_1\gamma+m) {\bar \Gamma^{(1)}}_\mu(q)\right\} \label{Wspur}
\end{eqnarray}
which involves the simply calculated trace and a function
$$C_d = \frac{1}{8\pi}\frac{1}{4e_{\k_1}e_{\pp_1}}S_{1\,+}^2(\k_1)
\sum_{l=0,2} |g_{l}(k_0,|\k|)|^2$$ containing the structure of the
deuteron. Performing the trace calculation we finally obtain the
expression for the hadron tensor
\begin{eqnarray}
W_{\mu\nu}= C_d\ \left( W^a_{\mu\nu} F_1^2(q^2) + W^b_{\mu\nu}
F_1(q^2)F_2(q^2) + W^c_{\mu\nu} F_2^2(q^2) \right)
\end{eqnarray}
with
\begin{eqnarray}
&&W^a_{\mu\nu} = 4\left[p_{1\,\mu} k_{1\,\nu}+k_{1\,\mu} p_{1\,\nu}
+(m^2-(p_1\cdot k_1))g_{\mu\nu}\right]
\label{Wparts}\\
&&W^b_{\mu\nu} =  2\left[k_{1\,\mu} q_{\nu}-q_{\mu} k_{1\,\nu}
-p_{1\,\mu} q_{\nu}+q_{\mu} p_{1\,\nu} \right]
\nonumber\\
&&W^c_{\mu\nu} =
\left[\left(-q^2 m^2-q^2(p_1\cdot k_1)+2(p_1\cdot q)(k_1\cdot q)\right)
g_{\mu\nu} + \left(m^2+(p_1\cdot k_1)\right)q_\mu q_\nu\nonumber\right.\\
&&-\left.\left((k_1\cdot q)(p_{1\,\mu} q_{\nu}+q_{\mu}p_{1\,\nu})
+(p_1\cdot q)(k_{1\,\mu} q_{\nu}+q_{\mu} k_{1\,\nu})
-q^2(p_{1\,\mu}
k_{1\,\nu}+k_{1\,\mu}p_{1\,\nu})\right)\right]/m^2.
\nonumber\end{eqnarray} Let us note here in the expressions
Eqs.~(\ref{Wspur},\ref{Wparts}) the four-vector $k_1$ has the
on-mass-shell form $k_1 = (e_{\k_1},\k_1)$ in differ with $k_1 =
(k_{10},\k_1)$ in the Fig.~\ref{kinematics}.

Thus we see that the factorization of the electrodisintegration
cross section exists both in nonrelativistic and relativistic
cases. The necessary conditions for this are the plane-wave
approximation for the final $np$-pair, the neutron in the deuteron
is supposed to be a spectator (the one-body type of the
interaction in the vertex ${\hat O}$) and only positive-energy
states for the deuteron are taking into account. As for the second
condition the type of one-body interaction does not play any role
but only spin-one-half particle is scattered. The third condition
means the $P$ waves in the deuteron (namely $^3P_1^{+-}$ and
$^1P_1^{+-}$) destroy the factorization.

\section{Results and Discussion}
\label{results}

We present here the results of the calculation of the deuteron
electrodisintegration cross section in the relativistic plane wave
impulse approximation with the separable Graz II rank III kernel
of interaction. In our calculations we follow the conditions of
real experiments and we distinguish eight sets of experimental
data. Let us mark these sets as $Saclay_I$, $Saclay_{II}$ (see
\cite{bernheim}, Table 3);  $Saclay_{III}$ (\cite{turck}, Table
1); $Bonn_{I}$, $Bonn_{II}$ (see \cite{breuker}, Table 3);
$Bonn_{III}$, $Bonn_{IV}$, $Bonn_{V}$ (see \cite{boden}, Tables
5,3,4, respectively). The kinematical conditions for all sets of
the experiment are shown in the Table~\ref{expsets}.

\begin{figure}[ht]
\begin{center}
\includegraphics[width=110mm]{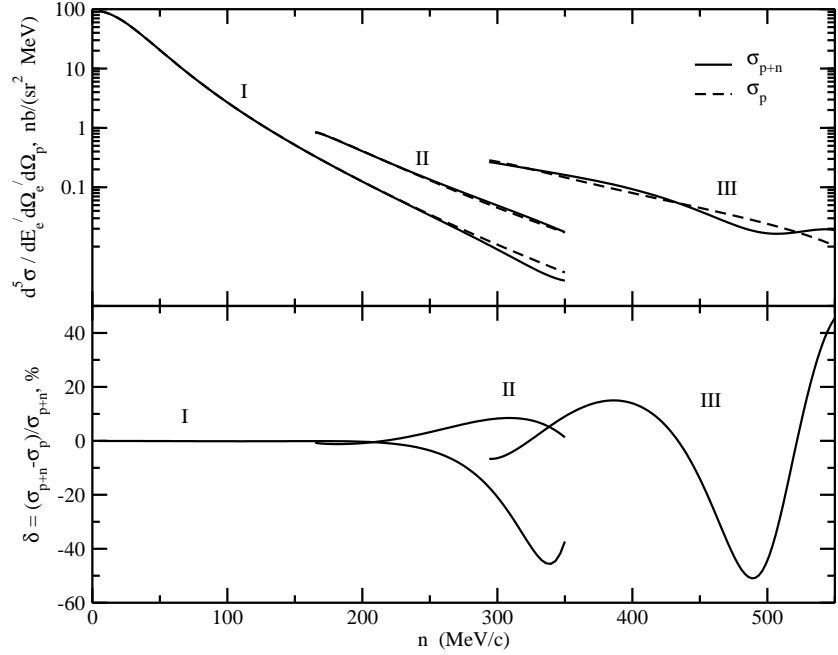}
\caption{{ The electrodisintegration cross section versus
the neutron momentum for three kinematics of the experiments at
Saclay. Solid and dashed lines correspond to the calculations with
and without neutron contribution (upper plot). Bottom plot shows
the relative neutron contribution in the corresponding
experimental regions. The experimental data regions were taken
from \protect\cite{bernheim}($Saclay_{I,II}$) and
\cite{turck}($Saclay_{III}$).}} \label{dkin_123c_bp_n}
\end{center}
\end{figure}

\begin{figure}[ht]
\begin{center}
\includegraphics[width=90mm]{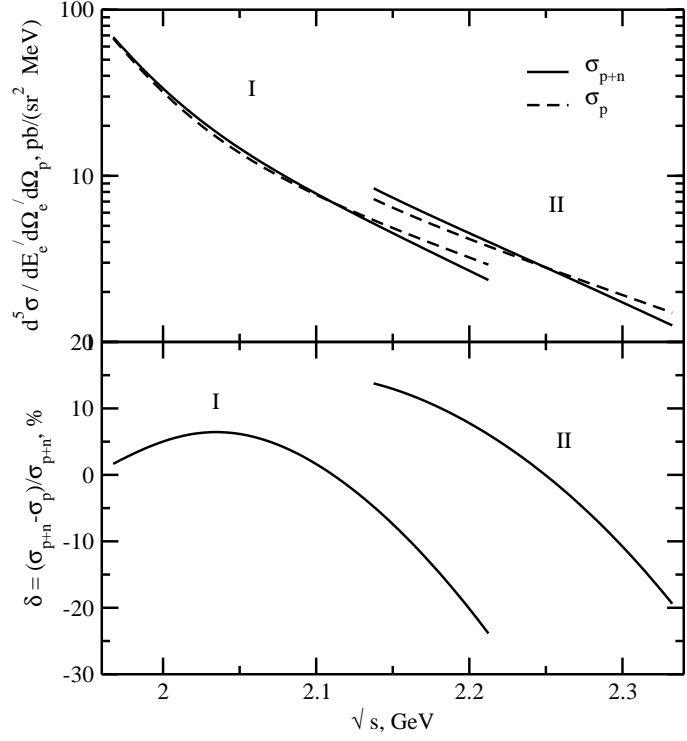}
\caption{{ The same as in previous figure but versus pair
invariant mass $\sqrt{s}$ for the kinematical conditions were
taken from \cite{breuker} ($Bonn_{I,II}$).}}
\label{dkin_s12c_bp_n}
\end{center}
\end{figure}

\begin{figure}[ht]
\begin{center}
\includegraphics[width=90mm]{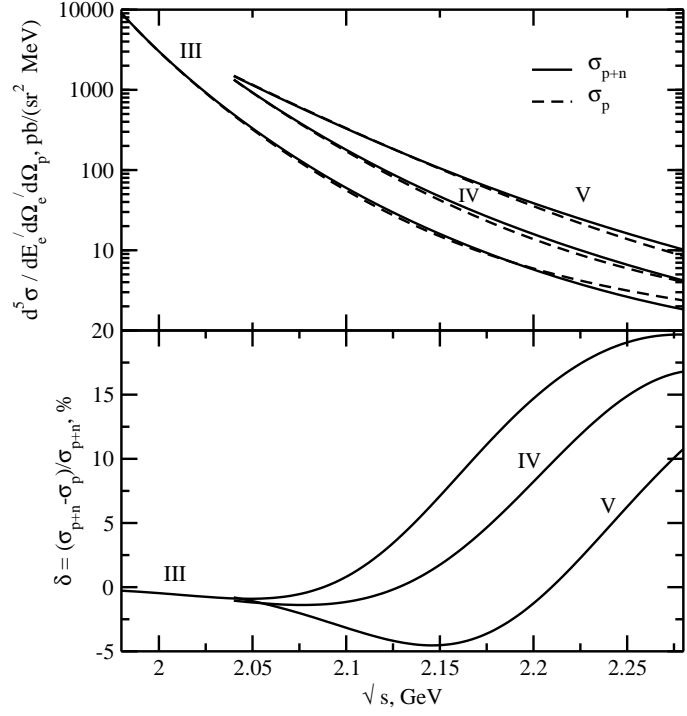}
\caption{{ The same as in previous figure. The kinematical
conditions were taken from \cite{boden}($Bonn_{III,IV,V}$).}}
\label{dkin_s3_123c_bp_n}
\end{center}
\end{figure}

\begin{figure}[ht]
\begin{center}
\includegraphics[width=110mm]{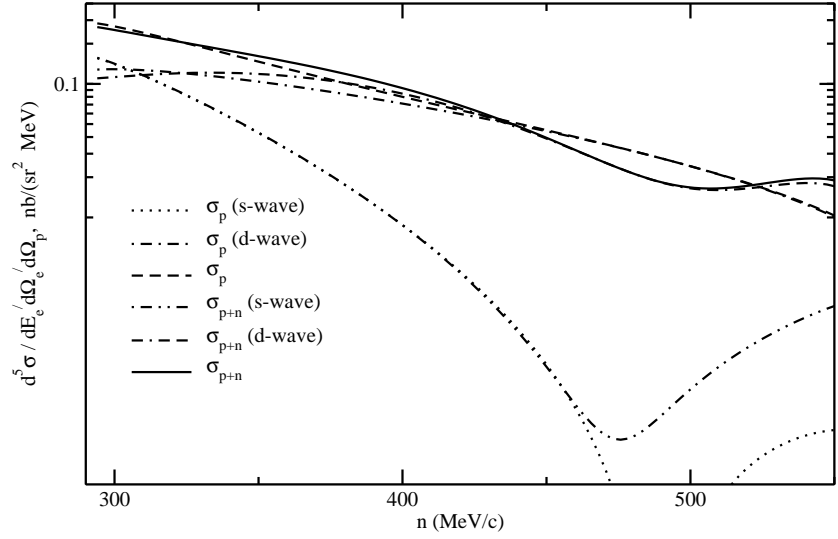}
\caption{{ The contributions of the spectator neutron versus
the outgoing neutron momentum to the electrodisintegration cross
section of the deuteron partial $S$-, $D$-states are shown for the
experimental sets from \cite{turck}($Saclay_{III}$).}}
\label{dkin_3c_bp}
\end{center}
\end{figure}

\begin{figure}[ht]
\begin{center}
\includegraphics[width=110mm]{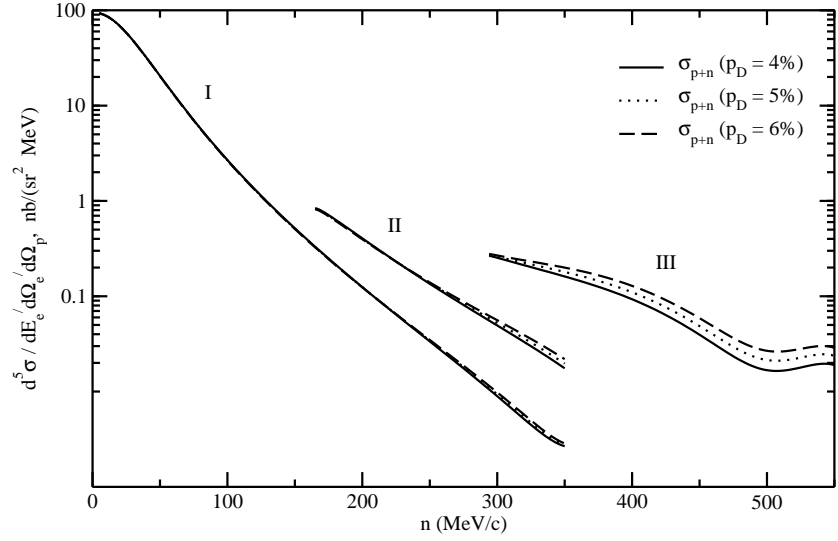}
\caption{{ The contributions of the deuteron partial D-state
to the electrodisintegration cross section versus neutron momenta
are shown for three sets of the experiments
\cite{bernheim}($Saclay_{I,II}$), \cite{turck} ($Saclay_{III}$).}}
\label{dkin_123c_pd}
\end{center}
\end{figure}

\begin{figure}[ht]
\begin{center}
\includegraphics[width=110mm]{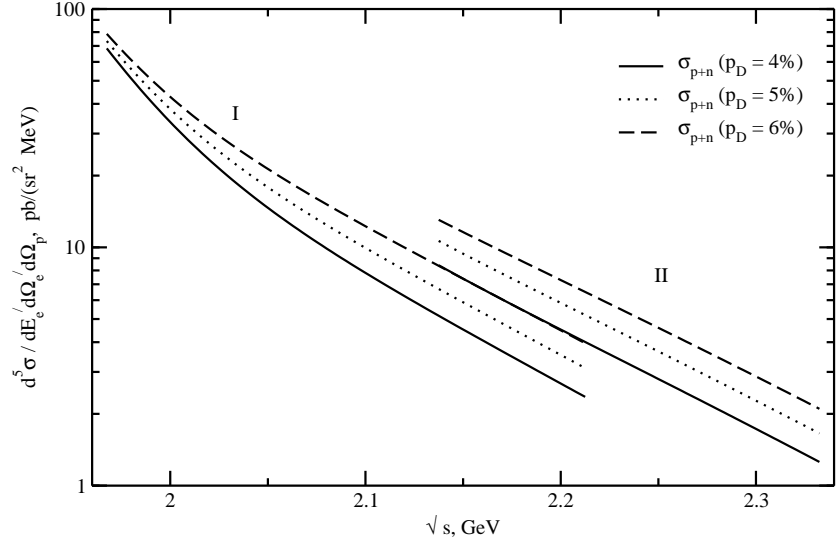}
\caption{{ The contributions of the deuteron partial D-state
to the electrodisintegration cross section versus pair invariant
mass $\sqrt{s}$ are shown for the conditions of the experiments
\cite{breuker} ($Bonn_{I,II}$).}} \label{dkin_s12c_pd}
\end{center}
\end{figure}

\begin{figure}[ht]
\begin{center}
\includegraphics[width=110mm]{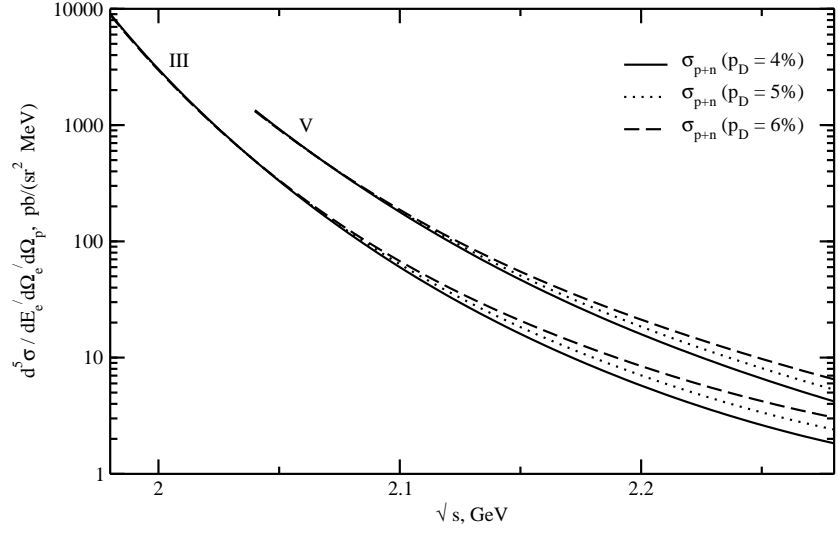}
\caption{{ The same as in the previous figure but for
\cite{boden} conditions ($Bonn_{III,V}$). We omitted curves for
the $Bonn_{IV}$ case because they are very close to $Bonn_{V}$.}}
\label{dkin_s3_123c_pd}
\end{center}
\end{figure}

\begin{figure}[ht]
\begin{center}
\includegraphics[width=110mm]{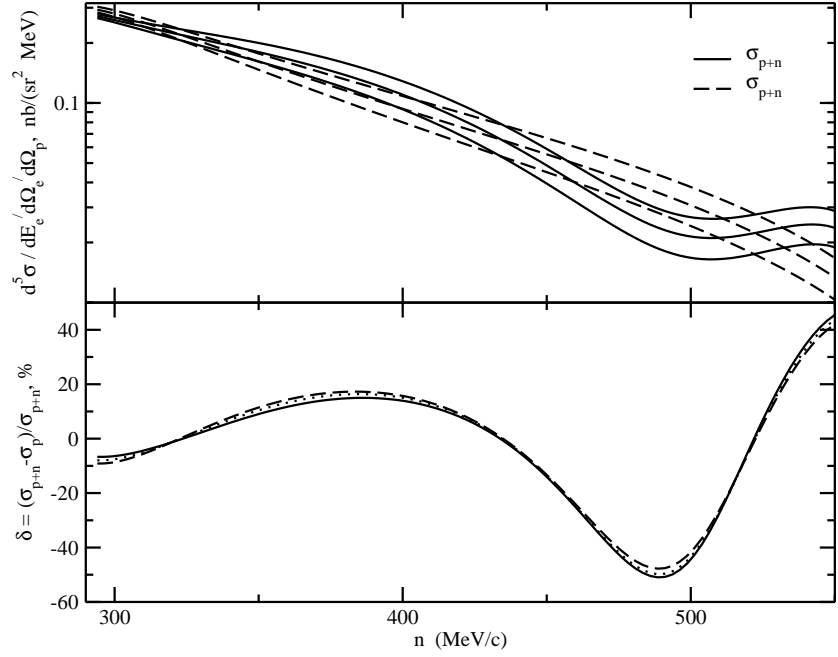}
\caption{{ The contribution of the spectator neutron versus
neutron momenta to the electrodisintegration cross section for
different deuteron $D$-states for conditions of the experiment
$Saclay_{III}$ \cite{turck}. In the first picture solid (dashed)
line stands for $p+n$- ($p$-) contribution with different
$D$-states in the deuteron: $p_D$ = 4\% for lower line and $p_D$ =
6\% for upper line.}} \label{dkin_3c_bpd}
\end{center}
\end{figure}

First of all we illustrated the influence of the spectator neutron
on the cross section (see Figs. \ref{dkin_123c_bp_n},
\ref{dkin_s12c_bp_n}, \ref{dkin_s3_123c_bp_n}). It is seen that it
increases with the increasing of the neutron momentum and reaches
50\% in the $Saclay_{III}$ kinematic range. One can see that the
cross section of the deuteron electrodisintegration
versus $\sqrt{s}$ changes not so strong, nevertheless  the
contribution of the spectator neutron is not negligible. Let us
note that this contribution changes sign in the $Saclay_{III}$
kinematic region (see Fig. \ref{dkin_123c_bp_n}). In order to
understand the origin of this behavior we present on the Fig.
\ref{dkin_3c_bp} partial contributions of the S- and D-states for
this  kinematical region versus neutron momenta.
We found that the D-state plays an important role and then it is
naturally to ask what happens if we change the magnitude of the
D-state. On the Figs. \ref{dkin_123c_pd}, \ref{dkin_s12c_pd},
\ref{dkin_s3_123c_pd}
we can see that for different magnitudes of the D-states the cross
section changes distinctly (especially for the kinematic region
\cite{turck}) but the ratio $\delta =
(\sigma_{p+n}-\sigma_p)/\sigma_{p+n}$ is not changed at all (see
Fig. \ref{dkin_3c_bpd}). It means that the difference is mainly
connected with the spectator neutron contribution. Thus we can
make a conclusion that the experimental data within the kinematics
from $Saclay_{III}$  \cite{turck} can supply the good test for
various models of NN interactions in the deuteron.

To check the influence of the relativistic effects we present the
results of the relativistic and nonrelativistic calculations for
various experimental conditions, see
figures~\ref{dkin_123b_rn}-\ref{dkin_s3_123b_rn}. It was shown
that relativistic effects play very important role even for small
transfer momenta $Q^2$ (see Table~\ref{expsets}). We can stress
that at the Fig.~\ref{dkin_s3_123b_rn} the difference mounts to
the order of value.
The next step is to take into account final state interactions,
$P$ -waves and to study the influence of nucleon ({\it on-shell}
and {\it off-shell}) form factors on the deuteron disintegration.

\section{Summary}
\label{sum}

In the presented paper we have considered the
electrodisintegration of the deuteron in the Bethe-Salpeter
approach. It is realized for a two-nucleon system by using the
multipole expansion with the spinor structure of two nucleons. The
separable ansatz for the interaction kernel has provided a
manageable system of linear homogeneous equations for deriving the
BS amplitude.

Then we have switched to
the using of the covariant revision of the Graz II separable
potential with the summation of several separable functions.

The reaction of the deuteron electrodisintegration served as a
testing ground for the method under investigation and helped to
outline both strong and weak points of the approach. The analysis
has proved the technique to be very promising, even if we find an
evident discrepancies with experimental data at this stage of
development. Several items can be suggested for the program of
further theoretical study. First of all it is necessary to take
into account the final state interaction for the $np$-pair. Then
we need to consider the negative-energy states for the BS
amplitude and calculate the contribution of $P$ waves to the
electrodisintegration (see, \cite{burov} and \cite{hamamoto}).
After that we will be able to calculate different asymmetries of
the $(ed\to enp)$ process which can give new qualitative
information about the structure of the deuteron.

\section{Acknowledgments}

We wish to thank our collaborators K. Yu.~Kazakov, A. V.~Shebeko,
S. Eh.~Shirmovsky, D. V. Shulga for their contribution to the
presented paper. We would like to thank Professor H.~Toki and
Professor D.~Blaschke for their interest to this work and fruitful
discussions.

\vspace*{3mm} The work is supported in part by the Russian
Foundation for Basic Research, grant No.05-02-17698a.

\newpage

\suppressfloats[hptb]

\newpage

\suppressfloats[hptb]

\newpage

\suppressfloats[hptb]

\newpage

\suppressfloats[hptb]

\newpage

\suppressfloats[hptb]

\newpage

\suppressfloats[hptb]

\newpage

\suppressfloats[hptb]

\newpage
\suppressfloats[hptb]

\newpage

\suppressfloats[hptb]

\newpage

\suppressfloats[hptb]

\newpage

\suppressfloats[hptb]

\newpage

\suppressfloats[hptb]
\begin{figure}[ht]
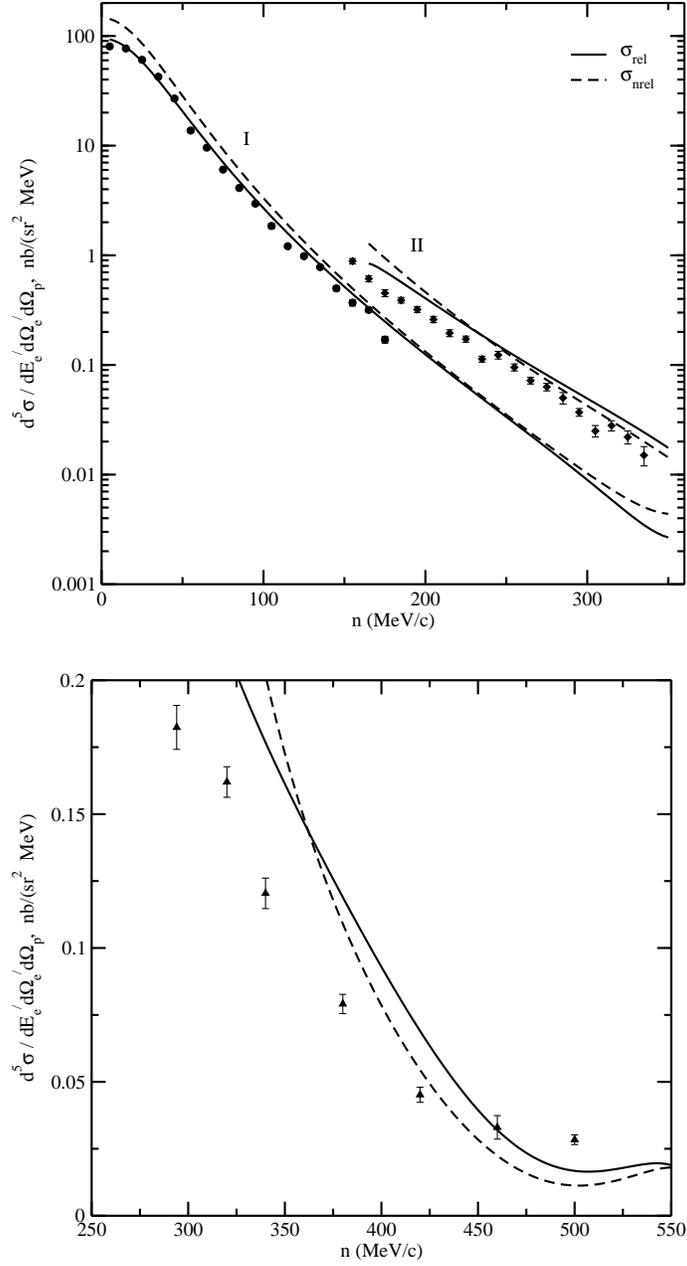

\begin{center}
\includegraphics[width=90mm]{figures/dkin_12b_rn.eps}\\[5mm]
\includegraphics[width=90mm]{figures/dkin_3b_rn.eps}
\caption{{ The relativistic and nonrelativistic
calculations for conditions of the experiments
$Saclay_{I,II}$ and $Saclay_{III}$. Experimental data are taken
from~\cite{bernheim} and \cite{turck}.}} \label{dkin_123b_rn}
\end{center}
\end{figure}

\newpage

\suppressfloats[hptb]
\begin{figure}[ht]
\begin{center}
\includegraphics[width=85mm]{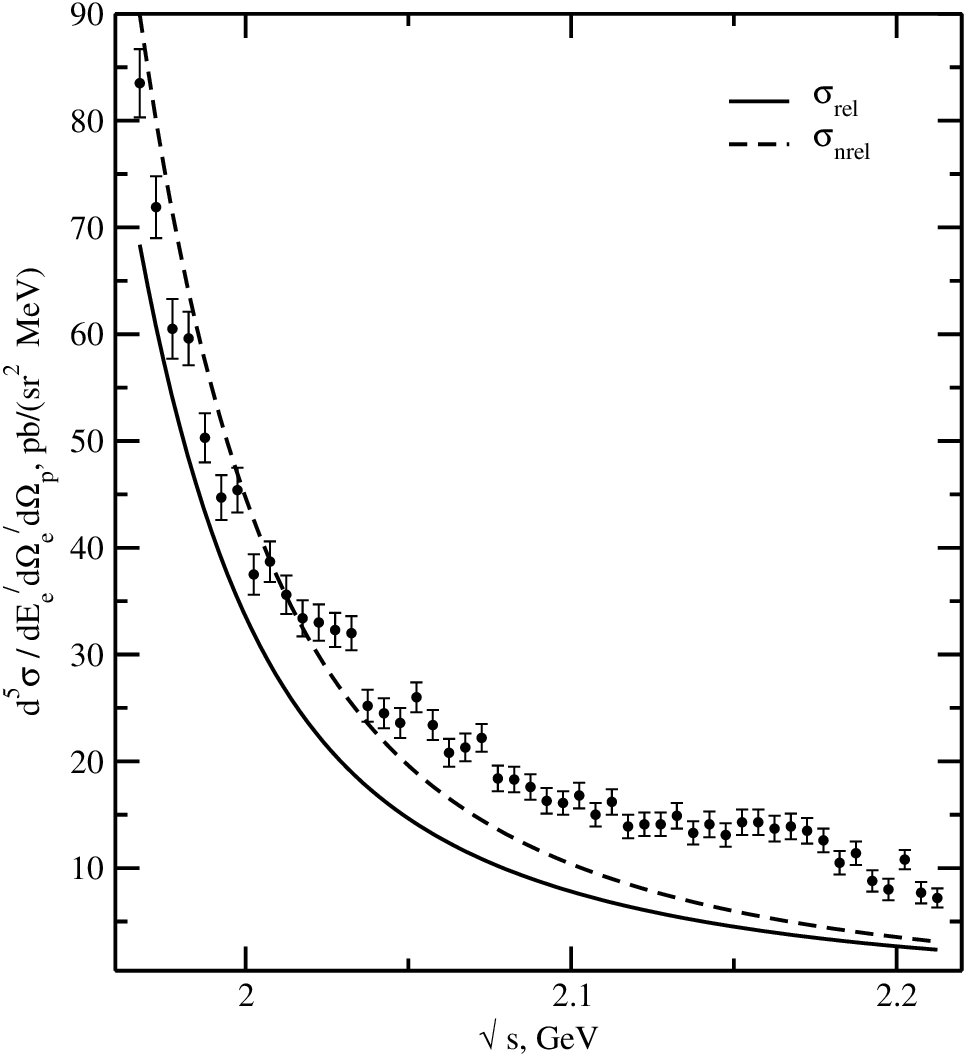}\\[5mm]
\includegraphics[width=85mm]{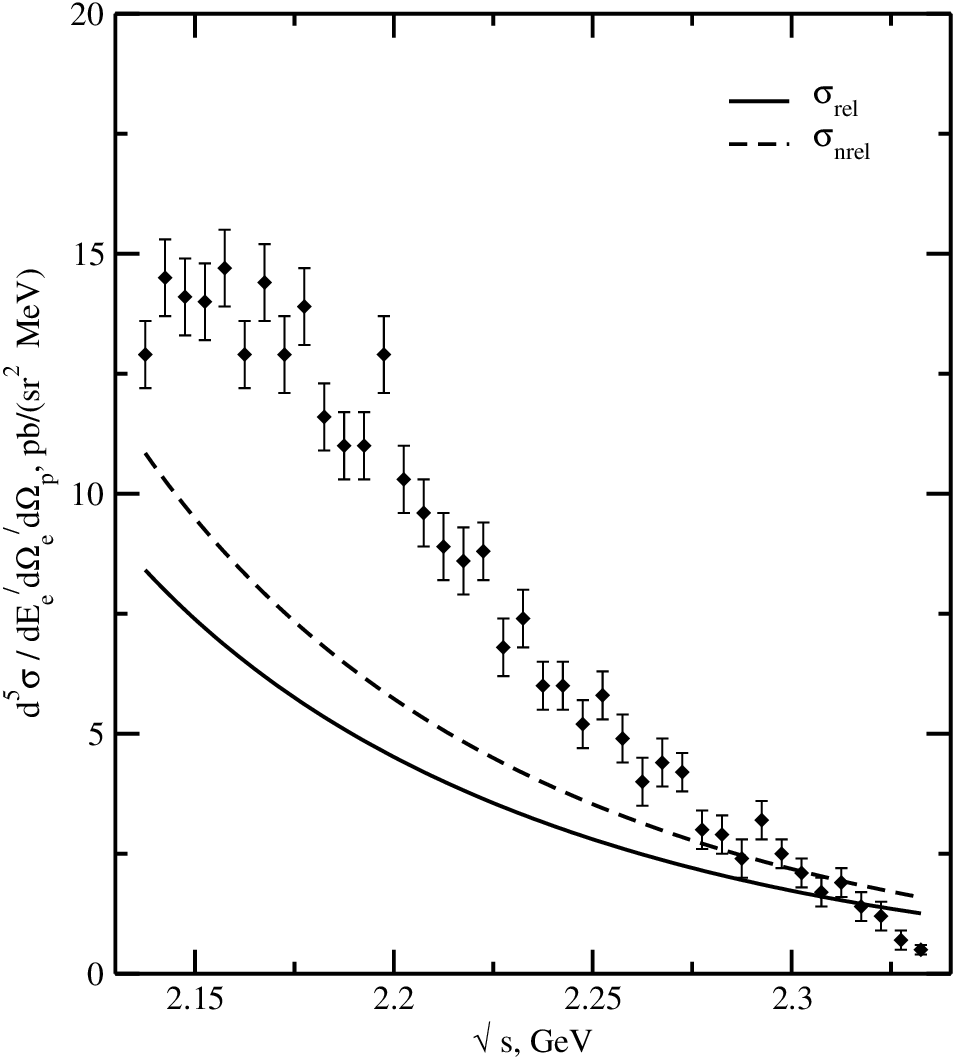}
\caption{{ The same as in Fig.~\ref{dkin_123b_rn}
for conditions of the experiments $Bonn_{I,II}$.
Experimental data are taken from~\cite{breuker}.}}
\label{dkin_s12b_rn}
\end{center}
\end{figure}

\newpage

\suppressfloats[hptb]
\begin{figure}[ht]
\begin{center}
\includegraphics[width=80mm]{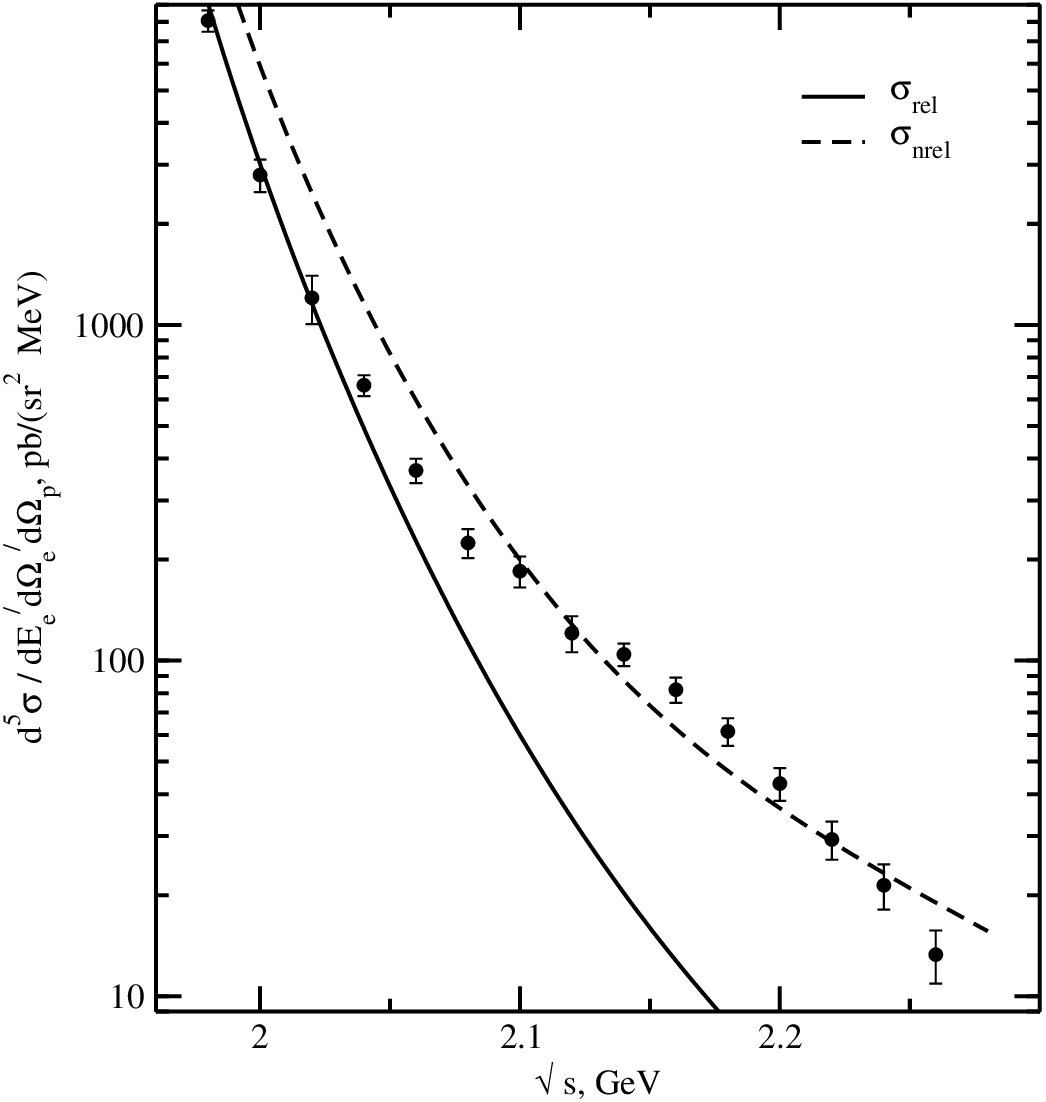}\\[5mm]
\includegraphics[width=80mm]{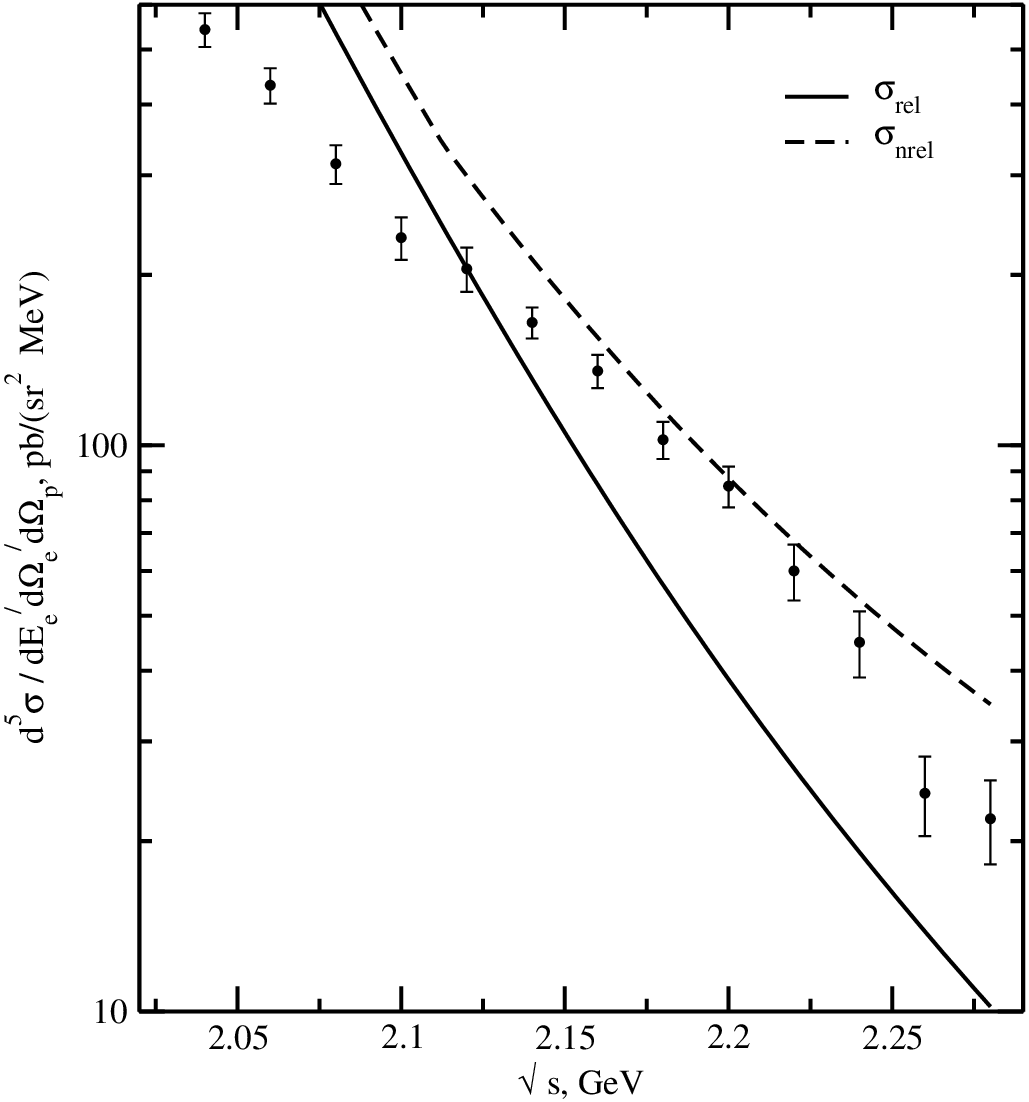}
\hskip 3mm
\includegraphics[width=80mm]{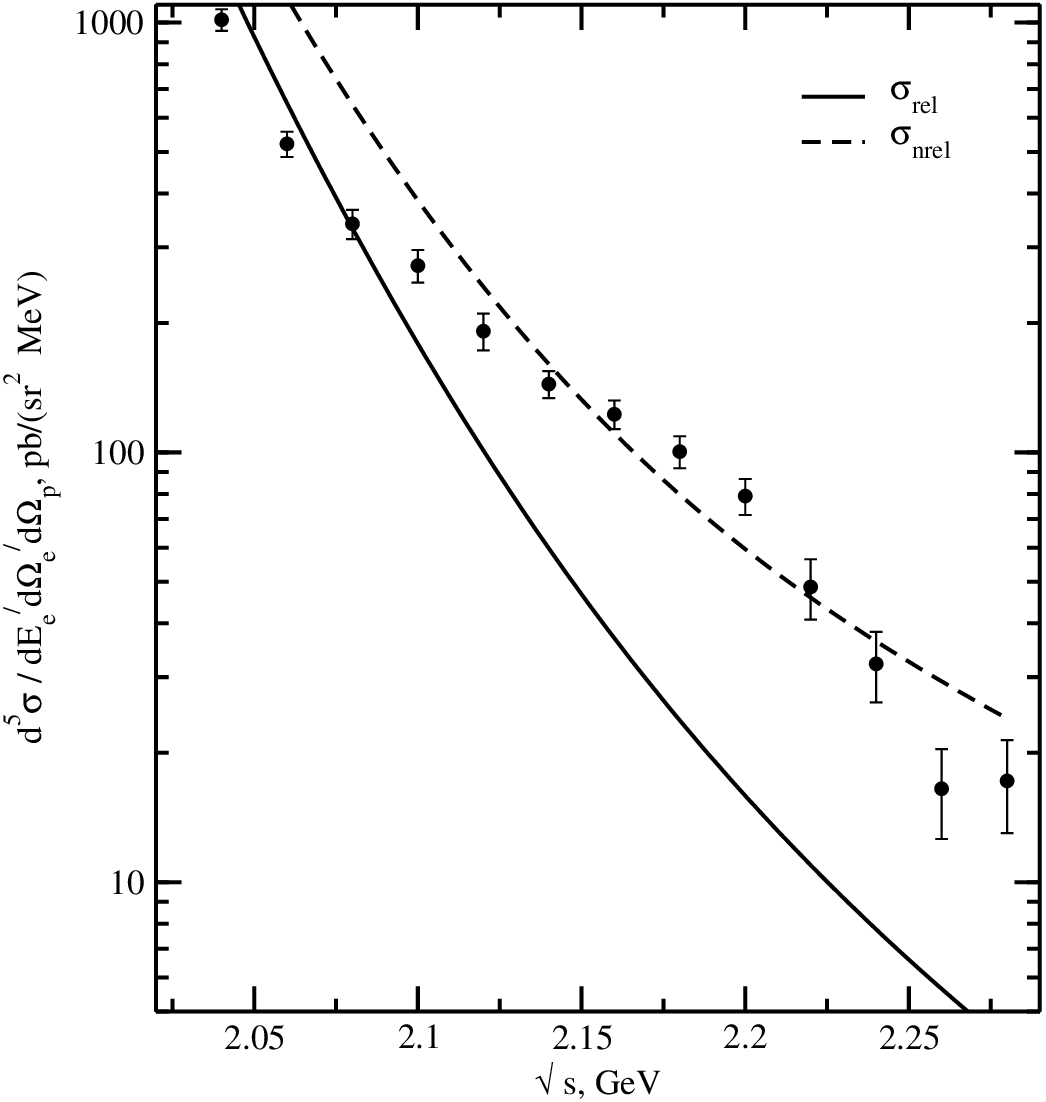}
\caption{{ The same as in Fig.~\ref{dkin_123b_rn} for
conditions of the experiments $Bonn_{III-V}$. Experimental data are
taken from~\cite{boden}.}} \label{dkin_s3_123b_rn}
\end{center}
\end{figure}

\newpage

\begin{table}[ht]
\begin{tabular}{|ll|c|c|c|c|c|c|c|c|}
\hline
            &         &  \sla   &  \slb   &  \slc   &  \boa   &  \bob   &  \boc   &  \bod   &  \boe  \\
\hline
\ee, $GeV$
            &         &  0.500  &  0.500  &  0.560  &  1.464  &  1.569  &   1.2   &    1.2  &   1.2  \\
\hline
\eepr, $GeV$
            & ${min}$ &  0.395  &  0.352  &  0.360  &  1.175  &  1.118  &  0.895  &  0.895  &  0.895 \\
            & ${max}$ &         &         &         &         &         &  0.800  &  0.800  &  0.800 \\
\hline
\thepr, ${}^{\circ}$
            &         &   59    &  44.4   &    25   &    21   &    21   &  20.15  &  20.15  &  20.15 \\
\hline
\pn, $GeV$
            & ${min}$ &  0.005  &  0.165  &  0.294  &  0.314  &  0.500  &  0.126  &  0.197  &  0.197 \\
            & ${max}$ &  0.350  &  0.350  &  0.550  &  0.660  &  0.773  &  0.564  &  0.423  &  0.488 \\
\hline
\thn, ${}^{\circ}$
            & ${min}$ & 101.81  &  172.07 & 153.01  &  60.53  &  74.60  & 142.32  & 155.72  & 165.36 \\
            & ${max}$ &  37.78  &  70.23  &  20.81  &  62.49  &  63.52  &  93.96  & 136.09  & 112.86 \\
\hline
\thqe, ${}^{\circ}$
            & ${min}$ &  48.79  &  44.74  &  33.06  &  61.94  &  45.57 &   59.56  &  51.52  &  51.25 \\
            & ${max}$ &         &         &         &  37.39  &  29.49 &   25.57  &  25.57  &  25.57 \\
\hline
\ppro, $GeV$
            & ${min}$ &  0.451  &  0.514  &  0.557  &  0.466  &  0.681  &  0.525  &  0.620  &  0.622 \\
            & ${max}$ &  0.276  &  0.403  &  0.306  &  0.664  &  0.791  &  0.834  &  0.929  &  0.889 \\
\hline
\thp, ${}^{\circ}$
            & ${min}$ &  0.622  &   2.54  &  13.86  &  35.82  &  45.12  &   8.42  &   7.52  &   4.47 \\
            & ${max}$ &  51.03  &  54.90  & 140.28  &  61.68  &  60.90  &  42.40  &  18.41  &  30.40 \\
\hline
\thpe, ${}^{\circ}$
            & ${min}$ &  49.41  &  47.28  &  46.92  &  97.77  &  90.68  &  68.00  &  44.00  &  56.00 \\
            & ${max}$ &  99.81  &  99.64  & 173.35  &  99.08  &  90.39  &         &         &        \\
\hline
\sqs, $GeV$
            & ${min}$ &  1.929  &  1.993  &  2.057  &  1.9675 &  2.1375 &  1.98   &  2.04   &  2.04  \\
            & ${max}$ &         &         &         &  2.2125 &  2.3325 &  2.28   &  2.28   &  2.28  \\
\hline
\sqsmm, $GeV$
            & ${min}$ &  0.051  &  0.115  &  0.176  &  0.090  &  0.260  &  0.101  &  0.161  &  0.161 \\
            & ${max}$ &         &         &         &  0.335  &  0.455  &  0.401  &  0.401  &  0.401 \\
\hline
$Q^2$, $GeV^2$
            & ${min}$ &  0.192  &  0.101  &  0.038  &  0.257  &  0.255  &  0.154  &  0.145  &  0.145 \\
            & ${max}$ &         &         &         &  0.206  &  0.209  &  0.106  &  0.106  &  0.106 \\
\hline
$\omega$, $GeV$
            & ${min}$ &  0.105  &  0.148  &  0.200  &  0.162  &  0.348  &  0.148  &  0.210  &  0.210 \\
            & ${max}$ &         &         &         &  0.422  &  0.568  &  0.476  &  0.476  &  0.476 \\
\hline
$q_z$, $GeV$
            & ${min}$ &  0.450  &  0.350  &  0.279  &  0.532  &  0.613  &  0.420  &  0.435  &  0.435 \\
            & ${max}$ &         &         &         &  0.620  &  0.729  &  0.577  &  0.577  &  0.577 \\
\hline
\end{tabular}
\caption{Kinematical conditions of the experiments under
consideration. All quantities are in laboratory system: \thqe\ is
an angle between the beam and the virtual photon, \pn - neutron
momentum, \thn  - angle between neutron and virtual photon, \ppro\
- proton momentum, \thp\ - angle between proton and virtual
photon, \thpe\ - angle between beam and proton. \sqsmm\ - kinetic
energy of the $np$-pair, $\omega$ and $q_z$ are the components of
the virtual photon four-momentum $q=(\omega,0,0,q_z)$. If value
$max$ is not stated it is equal to upper $min$. \label{expsets}}
\end{table}

\newpage
~~\\
{\bf Fig. 1:} One photon approximation.\\~~\\
{\bf Fig. 2:} Relativistic impulse approximation.\\~~\\
{\bf Fig. 3:} Plane wave approximation.\\~~\\
{\bf Fig. 4:} The electrodisintegration cross section versus the
neutron momentum for three kinematics of the experiments at
Saclay. Solid and dashed lines correspond to the calculations with
and without neutron contribution (upper plot). Bottom plot shows
the relative neutron contribution in the corresponding
experimental regions. The experimental data regions were taken
from \protect\cite{bernheim}($Saclay_{I,II}$) and
\cite{turck}($Saclay_{III}$).\\~~\\
{\bf Fig. 5} The same as in previous figure but versus pair
invariant mass $\sqrt{s}$ for the kinematical conditions were
taken from \cite{breuker} ($Bonn_{I,II}$).\\~~\\
{\bf Fig. 6:} The same as in previous figure. The kinematical
conditions were taken from \cite{boden}($Bonn_{III,IV,V}$).\\~~\\
{\bf Fig. 7:} The contributions of the spectator neutron versus
the outgoing neutron momentum to the electrodisintegration cross
section of the deuteron partial $S$-, $D$-states are shown for the
experimental sets from \cite{turck}($Saclay_{III}$).\\~~\\
{\bf Fig. 8:} The contributions of the deuteron partial D-state to
the electrodisintegration cross section versus neutron momenta are
shown for three sets of the experiments
\cite{bernheim}($Saclay_{I,II}$), \cite{turck} ($Saclay_{III}$).\\~~\\
{\bf Fig. 9:} The contributions of the deuteron partial D-state to
the electrodisintegration cross section versus pair invariant mass
$\sqrt{s}$ are shown for the conditions of the experiments
\cite{breuker} ($Bonn_{I,II}$).\\~~\\~~\\
{\bf Fig. 10:} The same as in the previous figure but for
\cite{boden} conditions ($Bonn_{III,V}$). We omitted curves for
the $Bonn_{IV}$ case because they are very close to $Bonn_{V}$.\\~~\\
{\bf Fig. 11:} The contribution of the spectator neutron versus
neutron momenta to the electrodisintegration cross section for
different deuteron $D$-states for conditions of the experiment
$Saclay_{III}$ \cite{turck}. In the first picture solid (dashed)
line stands for $p+n$- ($p$-) contribution with different
$D$-states in the deuteron: $p_D$ = 4\% for lower line and $p_D$ =
6\% for upper line.\\~~\\
{\bf Fig. 12:} The relativistic and nonrelativistic calculations
for conditions of the experiments $Saclay_{I,II}$ and
$Saclay_{III}$. Experimental data are taken from~\cite{bernheim}
and \cite{turck}.\\~~\\
{\bf Fig. 13:} The same as in Fig.~\ref{dkin_123b_rn} for
conditions of the experiments $Bonn_{I,II}$. Experimental data are
taken from~\cite{breuker}.\\~~\\
{\bf Fig. 14:} The same as in Fig.~\ref{dkin_123b_rn} for
conditions of the experiments $Bonn_{III-V}$. Experimental data
are taken from~\cite{boden}.\\~~\\
{\bf Table I:} Kinematical conditions of the experiments under
consideration. All quantities are in laboratory system: \thqe\ is
an angle between the beam and the virtual photon, \pn - neutron
momentum, \thn  - angle between neutron and virtual photon, \ppro\
- proton momentum, \thp\ - angle between proton and virtual
photon, \thpe\ - angle between beam and proton. \sqsmm\ - kinetic
energy of the $np$-pair, $\omega$ and $q_z$ are the components of
the virtual photon four-momentum $q=(\omega,0,0,q_z)$. If value
$max$ is not stated it is equal to upper $min$.


\newpage
\begin{thebibliography}{99}

\bibitem{crossaux}
M. Crossaux, Phys. Rev. {\bf 127}, 613 (1962).

\bibitem{arenh3}
T. Wilbois, G. Beck, H. Arenhovel, Few-Body Syst. {\bf 15}, 39 (1993).

\bibitem{arenh4}
G. Beck, T. Wilbois, H. Arenhovel, Few-Body Syst. {\bf 17}, 91 (1994).

\bibitem{gross1}
W. W. Buck, F. Gross, Phys. Rev. {\bf D20}, 2361 (1979).

\bibitem{gross}
V. Dmitrasinovic, F. Gross, Phys. Rev. {\bf C40}, 2479 (1989).

\bibitem{shebeko}
V. V. Kotlyar, Yu. P. Melnik, A. V. Shebeko,
Part. Nucl. {\bf 26}, 192 (1995).

\bibitem{gakh}
G. I. Gakh, A. P. Rekalo, Egle Tomasi-Gustafsson,
Ann. Phys. {\bf 319}, 150 (2005).

\bibitem{bernheim} 
M. Bernheim {\it et al.}, Nucl. Phys. {\bf A365}, 349 (1981).

\bibitem{turck} 
S. Turck-Chieze {\it et al.}, Phys. Lett. {\bf 142B}, 145 (1984).

\bibitem{breuker} 
H. Breuker {\it et al.}, Nucl. Phys. {\bf A455}, 641 (1986).

\bibitem{Bern}
M. Bernheim {\it et al.}, Phys. Rev. Lett. {\bf 46}, 402 (1981).

\bibitem{Auff} %
S. Auffret {\it et al.}, Phys. Rev. Lett. {\bf 55}, 1362 (1985).

\bibitem{boden} 
B. Boden {\it et al.}, Nucl. Phys. {\bf A549}, 471 (1992).

\bibitem{kiss}
T.S. Cheng, L.S. Kisslinger, Nucl. Phys. {\bf A457}, 602 (1986).

\bibitem{our-quarks}
V. V. Burov, S. M. Dorkin, V. N. Dostovalov, Z. Phys. A:
         Atoms and Nuclei {\bf 315}, 205 (1984);
V. V. Burov, V. K. Lukyanov,  Nucl. Phys. {\bf A463}, 263 (1987).

\bibitem{adam}
Jr. J. Adam, E.Truhlik, D. Adamova, Nucl. Phys. {\bf A492}, 556 (1989).

\bibitem{BS}
E. E. Salpeter, H. A. Bethe, Phys. Rev. {\bf 84}, 1232 (1951).

\bibitem{burov}
S. G. Bondarenko {\it et al.},
Prog. Part. Nucl. Phys. {\bf 48}, 449 (2002).

\bibitem{mandelstam}
S. Mandelstam, Proc. Roy. Soc. Lond. {\bf A233}, 248 (1955).

\bibitem{mathe}
L. Mathelitsch, W. Plessas, M. Schweiger, Phys. Rev. {\bf C26},
65-76 (1982).

\bibitem{hamamoto} S. G. Bondarenko {\it et.al.}, Part. and Nucl., Lett.,  {\bf 2}, 17 (2005).

\end{thebibliography}
\end{document}